\documentclass[aps,prl,twocolumn,groupedaddress,showpacs]{revtex4}
\usepackage{amsmath}
\usepackage{amsbsy}
\pdfoutput=1
\usepackage{graphicx}
\usepackage{upgreek}
\usepackage{amssymb}
\usepackage{epstopdf}
\usepackage{color}
\begin{document}
\title{Sub-diffraction Focussing using Quadratic Measure Eigenmodes}%

\author{Michael Mazilu}
\email{mm17@st-andrews.ac.uk}
%\homepage[]{Your web page}
%\thanks{}
%\altaffiliation{}
\affiliation{SUPA, School of Physics and Astronomy, University of St. Andrews, St. Andrews, KY16 9SS, UK}
\author{J\"org Baumgartl}%
%\email{jb211@st-andrews.ac.uk}
%\homepage[]{Your web page}
%\thanks{}
%\altaffiliation{}
\affiliation{SUPA, School of Physics and Astronomy, University of St. Andrews, St. Andrews, KY16 9SS, UK}
\author{Kishan Dholakia}%
%\email{kd1@st-andrews.ac.uk}
%\homepage[]{Your web page}
%\thanks{}
%\altaffiliation{}
\affiliation{SUPA, School of Physics and Astronomy, University of St. Andrews, St. Andrews, KY16 9SS, UK}

\date{\today}

\begin{abstract}
For over a century diffraction theory has been thought to limit the resolution of focusing and imaging in the optical domain. The size of the smallest spot achievable is inversely proportional to the range of spatial wavevectors available. Here, we show that it is possible to locally beat the diffraction limit at the expense of efficiency. The method is based on the linearity of Maxwell's equations and that the interaction between light and its surroundings may be considered quadratic in nature with respect to the electromagnetic fields. We represent the intensity and spot size as a quadratic measure with associated eigenmodes. Using a dynamic diffractive optical element, we demonstrate optical focussing to an area 4 times smaller than the diffraction limit. The generic method may be applied to numerous physical phenomena relating to linear and measurable properties of the electromagnetic field that can be expressed in a quadratic form.
\end{abstract}

\maketitle

\paragraph{Introduction}
The diffraction limit may be seen to originate from the Heisenberg uncertainty principle \cite{Heisenberg1927,Stelzer2000}. The concept of super-resolution to beat the so termed ``diffraction limit'' is typically associated with the retention of evanescent fields which are needed to retain all $k$-space components to realise ``perfect imaging'' \cite{Betzig1991,Li2003,Merlin2007}. The concept of a superlens \cite{Pendry2000} is typically based on the recovery of evanescent information by judicious use of suitable metamaterials - artificial materials exhibiting a negative refractive index \cite{Smith2004}. However, it would be advantageous to explore methods of overcoming the diffraction limit in the far field, even if over a restricted range, as this is amenable to the vast majority of optical systems.  Intriguingly, recent work on band-limited functions oscillating faster than the highest Fourier components of which they are composed, so termed superoscillations, shows that there may be routes to sub-diffraction imaging in the optical far-field without the need to retain rapidly decaying evanescent waves \cite{Zheludev2008,Berry2006,Huang2007,Huang2009}. However, with superoscillations, we achieve sub-diffraction features at the expense of having most of the energy contained in low frequency Fourier features (side-bands) that have many orders of magnitude higher amplitude than any sub-diffractive feature we may wish to utilise.  Speckled fields, originating from strong (diffuser-like) phase aberrations, can also deliver random positioned sub-diffraction patterns or hot spots. Their usefulness is limited due to their randomness and the lack of control of the distribution of energy between the speckles. It thus remains an open question as to how to potentially realise sub-diffraction focusing and imaging in a generic manner that is applicable as this would open up a wealth of new scientific directions.

Here we address the problem of realising the intensity of an optical field such that we produce a focussed spot smaller that the diffraction limit dictated by the optical system employed. The most intuitive way to describe our approach is as a superposition of incident fields that is optimised to achieve this smallest spot. Naturally, multiple techniques can be employed for this optimisation process ranging from a genetic algorithm to a steepest descent method. The major challenge encountered in any such optimisation and engineering of optical intensities is the fact that electromagnetic waves interfere. Indeed, the intensity profile arising from the superposition of multiple fields is a complex interference pattern depending not only upon the intensities of the considered waves but also upon their relative phase fronts. This interference pattern not only makes the search for an optimum beam problematic but crucially renders the superposition found unreliable as the different algorithms may converge on local minima which are unstable with respect to the different initial parameters in the problem. In contrast, our proposed method yields a unique solution to the problem and directly yields sub-diffraction optical features in the far field of our designated optical system. As might be expected, our method gives insights into the area of superoscillations which we shall return to in the conclusions.

Our method is based on two fundamental properties of the electromagnetic field and its interactions. Firstly, the approach relies on the linearity of the electromagnetic fields i.e. that the sum of two solutions of Maxwell's equations is itself a solution of them. As we consider free space propagation, this criteria is satisfied. The second property relates to the interaction of the electromagnetic field with its environment. All such interactions can be written in the form of quadratic expressions with respect to the electric and magnetic fields. Examples include the energy density, the energy flow and Maxwell's stress tensor. This allows us to designate appropriate ``quadratic measure'' eigenmodes to various parameters (\emph{e.g.} spot size) and subsequently ascertain the minimum eigenvalue which, in the case of a spot size operator yields a sub-diffraction optical feature. We first present the theory underlying our approach and then demonstrate experimentally the applicability of our method to achieve sub-diffraction focussing.

\paragraph{Quadratic measure eigenmodes (QME)} Our method assumes monochromatic solutions of free space Maxwell's equations, with $\mathbf{E}$ and $\mathbf{H}$ the electric and magnetic vector fields and with $\epsilon_{0}$ and $\mu_{0}$ the vacuum permittivity and permeability. These solutions can be written in an integral form linking the electromagnetic fields on the surface $A$ with the fields at any position $\mathbf r$
\begin{equation}
\label{diffract}
\mathcal{F}_u(\mathbf r) =\int_{A}\mathbf{P}_{uv}(\mathbf{r},\mathbf{r}')\mathcal{F}_v(\mathbf{r}')dS'
\end{equation}
where $\sqrt{2} \mathcal F=(\sqrt{\epsilon_0}\mathbf E,\sqrt{\mu_0}\mathbf H)$ is a shorthand for the two electromagnetic fields having six $\mathcal F_u$ scalar components. The integration kernel $\mathbf P_{uv}$ corresponds to a propagation operator giving rise to different vector diffraction integrals such as  Huygens, Kirchhoff and Stratton-Chu. Crucially all linear and measurable properties of the electromagnetic field can be expressed as quadratic forms of the local vector fields and are therefore termed \emph{quadratic measures}. For instance, the time averaged energy density of the field is proportional to $\mathcal{F}^{\ast}\cdot\mathcal{F}=1/2(\epsilon_0\mathbf{E}^{\ast}\cdot\mathbf{E}+\mu_0\mathbf{H}^{\ast}\cdot\mathbf{H})$ while the energy flux to $1/2 (\mathbf{E}^{\ast}\times\mathbf{H}+\mathbf{E}\times\mathbf{H}^{\ast})$. The asterisk $^{\ast}$ stands for the complex conjugate. Integrating the first quantity over a volume determines the total electromagnetic energy in this volume while integrating the normal energy flux across a surface the intensity of the light field incident on this surface. All the quadratic measures can be represented in a compact way by considering the integral
\begin{equation}
\label{QM:genl}
M_\kappa=\int_V\mathcal{F}_u^{\ast}\boldsymbol{\kappa}_{uv}\mathcal{F}_v d \mathbf{r} =\langle\mathcal{F}|\boldsymbol{\kappa}|\mathcal{F}\rangle _V 
\end{equation}
where the kernel $ \kappa_{uv} = \kappa^{\dagger}_{vu} $ is Hermitian where $\dagger$ the adjoint operator including boundary effects for finite volumes. Table \ref{tab} enumerates some operators associated to common quadratic measures. The integrand part of all these quadratic measures correspond to the conserving densities which together with the associated currents is Lorentz invariant \cite{Mazilu2009}. The volume, over which the integral is taken, does not need to be the whole space and can be a region of space, a surface, a curve or simply multiple points. To account for this general integration volume, we broadly term it the region of interest (ROI) in the following. Finally, using the general definition of the quadratic measure it is possible to define a Hilbert sub-space, over the solutions of Maxwell's equations, with the energy operator (EO) defining the inner product. Further, any general quadratic measure defined by (\ref{QM:genl}) can be represented in this Hilbert space by means of its spectrum of eigenvalues and eigenfunctions defined by 
$\lambda \mathcal F_u=\kappa_{uv}  \mathcal F_v$. Depending on the operator $\kappa_{uv}$, the eigenvalues $\lambda$ form a continuous or discrete real valued spectra which can be ordered. This gives direct access to the solution of Maxwell's equations with the largest or smallest measure. The eigenfunctions are orthogonal to each other ensuring simultaneous linearity in both field and measure. In the following, we study the case of different quadratic measure operators and their spectral decomposition into modes which we term the \emph{quadratic measure eigenmodes (QME)}. The convention for operator labelling we adopt is to use the shorthand QME followed by a colon and a shorthand of the operator name.

\begin{table}[htb]
\begin{tabular}{|c|c|}
\hline
Operator  & $2\mathcal F_u^{\ast} \boldsymbol \kappa_{uv}  \mathcal F_v $ \\ \hline
EO & $  \epsilon_0 \mathbf{E}^*\cdot\mathbf{E}+\mu_0 \mathbf{H}^*\cdot\mathbf{H}$ \\
IO  &$( \mathbf{E}^*\times\mathbf{H}+\mathbf{E}\times\mathbf{H}^*) \cdot \mathbf u_k$ \\
SSO  & $\mathbf r^2  ( \mathbf{E}^*\times\mathbf{H}+\mathbf{E}\times\mathbf{H}^*) \cdot \mathbf u_k$ \\
LMO  & $  \epsilon_0 \mathbf{E}^*\cdot(i\partial_k)\mathbf{E}+\mu_0 \mathbf{H}^*\cdot(i\partial_k)\mathbf{H}  $ \\
OAMO  & $ \epsilon_0 \mathbf{E}^*\cdot(i\mathbf r\times \nabla)_k\mathbf{E}+\mu_0 \mathbf{H}^*\cdot (i\mathbf r\times \nabla)_k\mathbf{H} $  \\
CSO  &$i( \mathbf{E}^*\cdot\mathbf{H}-\mathbf{H}^*\cdot\mathbf{E})$ \\
\hline
\end{tabular}
\caption{Common quadratic measure operators including the energy operator (EO), intensity operator (IO), spot size operator (SSO), linear momentum operator (LMO), orbital angular momentum operator (OAMO) and circular spin operator (CSO). The vector operators include the subscript $k$ indicating the different coordinates and $\mathbf u_k$ the associated unit vectors.}
\label{tab}
\end{table}

\paragraph{Practical example: smallest focal spot}
\label{s:pracex}
In our practical example we utilised the intensity operator \emph{QME:IO} and the spot size operator \emph{QME:SSO} defined in the following to engineer the size of a laser focus. The QME:IO measures the electromagnetic energy flow across a surface $A$: 
\begin{equation}
\label{eq:QMEIO2}
m^{(0)}=\frac{1}{2}\int_{\textrm{ROI}} ( \mathbf{E}^*\times\mathbf{H}+\mathbf{E}\times\mathbf{H}^*) \cdot \mathbf n dS
\end{equation}
where $\mathbf n$ is normal to the surface of interest. The eigenvector decomposition of this operator can be used for example to maximise the optical throughput through a pinhole or to minimise the intensity in dark spots. Considering a closed surface  surrounding an absorbing particle, the QME of the IO give access to the field that either maximises or minimises the absorption of this particle. The definition of the QME:SSO is based on the concept of determining the spot size of a laser beam by measuring, keeping the total intensity constant, the second order momentum of its intensity distribution. 
\begin{equation}
\label{eq:QMESSO2}
m^{(2)}=\frac{1}{2}\int_{\textrm{ROI}}  |\mathbf r-\mathbf r_0|^2  ( \mathbf{E}^*\times\mathbf{H}+\mathbf{E}\times\mathbf{H}^*) \cdot \mathbf n dS
\end{equation}
where $\mathbf r$ is the position vector and $\mathbf r_0$ the centre of the beam. The eigenvalues of this operator measure the spread of the beam with respect to its centre and the smallest eigenvalue defines the smallest spot achievable in the ROI. In the following, we define $\sigma=\sqrt{m^{(2)}}$ as the width of this spot.

For the experimental determination of both the QME:IO and QME:SSO, we considered a pair of an initial and a target plane located at the propagation distances $z=z_{I}$ and $z=z_{T}$ and connected through a linear optical system. Crucially, a superposition of fields $E(x,y,z_{\textrm{I}})=\sum_{u=1}^{N_{u}}a_{u}E_{u}(x,y,z_{\textrm{I}})$ ($a_{u}\in\mathbb{C}$, $N_{u}\in\mathbb{N}$) in the initial plane is rendered into a superposition of the respective propagated fields $E(x,y,z_{\textrm{T}})=\sum_{u=1}^{N_{u}}a_{u}E_{u}(x,y,z_{\textrm{T}})$ characterised by the same set of coefficients $a_{u}$ due to linearity of the optical system. Based on this superposition approach and the QME:IO as defined in Eq. \eqref{eq:QMEIO2}, the intensity in the target plane can be represented as
$m^{(0)}=\mathbf{a}^{\ast}{\mathbf{M}}^{(0)}\mathbf{a}$. $\mathbf{M}^{(0)}$ is a $N\times{}N$ matrix with the elements given by the overlap integrals 
\begin{equation}
\label{eq:QMEIO1}
M_{uv}^{(0)}=\int_{\textrm{ROI}}E^*_{u}(x,y,z_{\textrm{T}})E_{v}(x,y,z_{\textrm{T}})dS.
\end{equation}
This matrix is equivalent to the QME:IO on the Hilbert subspace defined by the fields ${E}_{u}(x,y,z_{\textrm{T}})$. $\mathbf{M}^{(0)}$ is Hermitian and positive-definite which implies that its eigenvalues $\lambda_{k}^{(0)}$ ($k=1\dots{}N_{u}$) are real and positive and the eigenvectors $\mathbf{v}_{k}^{(0)}$ are mutually orthogonal. Accordingly the largest eigenvalue $\lambda_{\max}^{(0)}=\max(\lambda_{k}^{(0)})$ and the associated eigenvector $\mathbf{v}_{\max}^{(0)}$ will deliver the superposition $E_{\max}(x,y,z)=\sum_{u=1}^{N_{u}}v_{\max,u}^{(0)}E_{u}(x,y,z)$ ($z=z_{\textrm{I}}$ and $z=z_{\textrm{T}}$ due to linearity) which maximizes the intensity within the ROI. Similar to the QME:IO, the QME:SSO as defined in Eq. \eqref{eq:QMESSO2} can be written as $m^{(2)}=\mathbf{b}^{\ast}{\mathbf{M}}^{(2)}\mathbf{b}$ where ${\mathbf{M}}^{(2)}$ must be represented in the intensity normalised base $\widetilde{E}_{k}(x,y,z_{\textrm{T}})=\sum_{u=1}^{N_{u}}(v_{k,u}^{(0)}/\lambda_{k}^{(0)})E_{u}(x,y,z_{\textrm{T}})$. $\mathbf{M}^{(2)}$ is a $N\times{}N$ matrix with the elements given by 
\begin{equation}
\label{eq:QMESSO1} 
M_{uv}^{(2)}=\int_{A}|\mathbf{r}-\mathbf{r}_{0}|^2\widetilde{E}_{u}^{\ast}(x,y,z_{\textrm{T}})\widetilde{E}_{v}(x,y,z_{\textrm{T}})dS.
\end{equation}
We denote the eigenvalues of ${\mathbf{M}}^{(2)}$ as $\lambda_{k}^{(2)}$ and the eigenvectors as $\mathbf{v}_{k}^{(2)}$. The eigenvector associated with the smallest eigenvalue corresponds to the smallest spot achievable within the ROI through the linear superposition of the $N_{u}$ fields $E_{u}(x,y,z_{\textrm{I}})$ considered initially.

\paragraph{Experiment}
Our experiments are based on the expressions \eqref{eq:QMEIO1} and \eqref{eq:QMESSO1} which allowed us to determine the superposition coefficients for the smallest spot from a set of test electric fields $E_{u}(x,y,z_{\textrm{T}})$ measured in the target plane. We used an expanded HeNe laser beam ($P=4\ \textrm{mW}$, $\lambda=633\ \textrm{nm}$) to illuminate the chip of a phase-only spatial light modulator (SLM, type Hamamatsu LCOS X10468-06, $800\ \textrm{pixel} \times600\ \textrm{pixel})$ operating in the standard first order configuration \cite{DiLeonardo2007}. Note that additional amplitude modulation can be achieved in a straightforward manner but is beyond the scope of our proof-of-principle study. Without lack of generality we have chosen the Zernike polynomials $Z_{n}^{m}(x,y)$ $(m,n\in\mathbb{N})$ \cite{Wang1980} to modulate the beam phase that is our test fields exhibited a phase behavior according to $E_{u}(x,y,z_{\textrm{I}})\propto\exp(i\cdot{}Z_{n}^{m}(x,y))$ where the index $u$ enumerates the different combinations $(n,m)$. The modulated beam was subsequently propagated through a spherical lens (focal width $f=1\ \textrm{m}$), the linear optical system, and then detected with a CCD camera (Basler pilot piA640-210gm). Since the CCD camera only detected intensities we applied the well-known lock-in technique to the optical domain as described in detail in the Supplementary information. In brief, a reference Gaussian beam, whose phase was oscillated in time using the SLM, was interfered with the test field in the target plane in order to determine both amplitude $A_{u}(x,y,z_{T})$ and phase $\phi_{u}(x,y,z_{T})$ of the test field in the target plane. The respective reference field parameters $A_{R}(x,y,z_{\textrm{T}})$ and $\phi_{R}(x,y,z_{\textrm{T}})$ were independently determined using self-interference and an approximate evaluation of the phase gradient field including subsequent numerical integration. Both the QME:IO and the QME:SSO were finally constructed from the measured parameters $A_{u}(x,y,z_{T})$ and $\phi_{u}(x,y,z_{T})$ according to Eqs. \eqref{eq:QMEIO1} and \eqref{eq:QMESSO1}.
Dedicated Labview and Matlab software allowed us to record a set of $N_{u}$ test fields  $E_{u}(x,y,z_{\textrm{T}})$ (typically $N_{u}=231$ corresponding to the Zernike polynomials up to order $n=20$) at a rate of $50\ \textrm{Hertz}$. Each test field required a 48 point temporal phase scan. Numerical evaluation of the QME:IO and QME:SSO finally delivered the required superposition $E(x,y,z_{\textrm{I}})=A(x,y,z_{\textrm{I}})e^{i\phi(x,y,z_{\textrm{I}})}=\sum_{i=1}^{N_u}v_{\min,u}^{(2)}E_{u}(x,y,z_{\textrm{I}})$ which was encoded onto the SLM. The final superposition required simultaneous modulation of both amplitude and phase of the laser beam incident onto the SLM which we have encoded to our phase-only SLM using the approximation $Ae^{i\phi}\approx{}e^{iA\phi}$ as described in detail elsewhere \cite{Davis1999}. Crucially, the QME:SSO was determined for decreasing size of the target ROI which allowed us to squeeze the laser spot size below the diffraction limit as shown in the following. 

\begin{figure}[bt]
\centering\includegraphics[width=8cm]{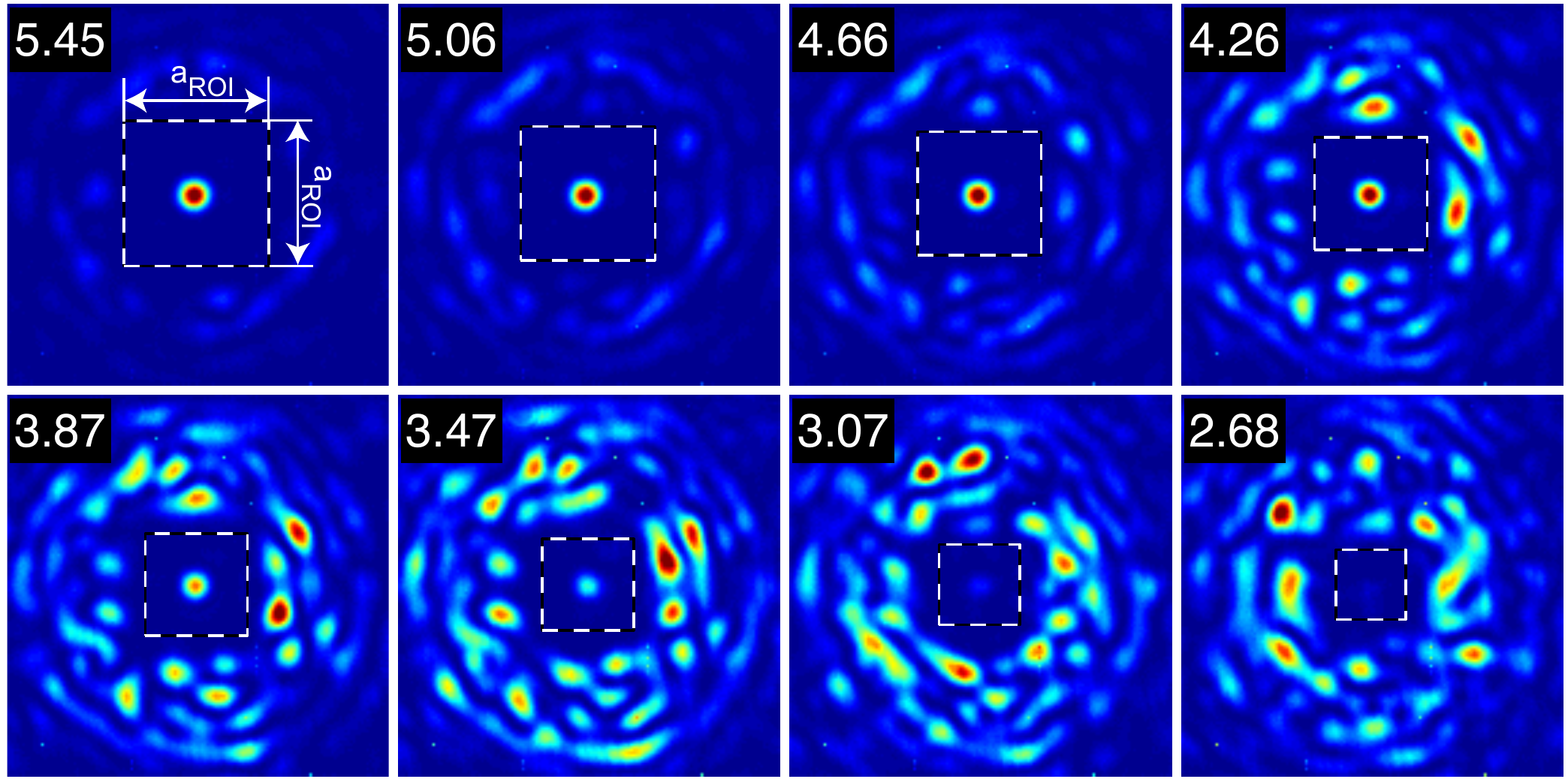}
\caption{2D intensity profiles recorded in the target plane for decreasing ROI size.  The lateral size of the profile plots is $1.04\ \textrm{mm}\times1.04\ \textrm{mm}$. The ROI is indicated by black and white dashed square. Numbers in the left top corner of the profiles indicate the ROI size in units of the Airy disk size $\sigma_{\textrm{ROI}}=63\ \upmu\textrm{m}$. Blue to red color indicates low to high intensity.}
\label{fig:col2Dexp}
\end{figure}

\paragraph{Results and discussion}
Figure \ref{fig:col2Dexp} shows a set of intensity profiles $I(x,y,z_{\textrm{T}})$ as obtained after encoding the final superposition of test fields for different target ROI sizes. The ROI side length $a_{\textrm{ROI}}$ is indicated in the profile's left top corner in units of the Airy disk size $\sigma_{\textrm{Airy-disk}}$ which was $\sigma_{\textrm{Airy-disk}}=63\ \upmu\textrm{m}$ given the laser wavelength $\lambda$ and the numerical aperture $\textrm{NA}=0.005$ of our optical setup. Crucially, the intensity profiles reveal a central spot whose size is decreasing when the ROI size is reduced. This is balanced by a redistribution of intensity into the area outside of the ROI. Interestingly, the applied procedure not only aims to achieve the smallest spot size possible for a given set of test fields but also clearly aims to keep the redistributed intensity entirely outside the square shaped ROI. The redistributed intensity starts to evolve at $\sigma/\sigma_{\textrm{Airy-disk}}\approx6$ (data not shown) and becomes predominant for $\sigma/\sigma_{\textrm{Airy-disk}}<4$. 

\begin{figure}[tb]
\centering\includegraphics[width=8cm]{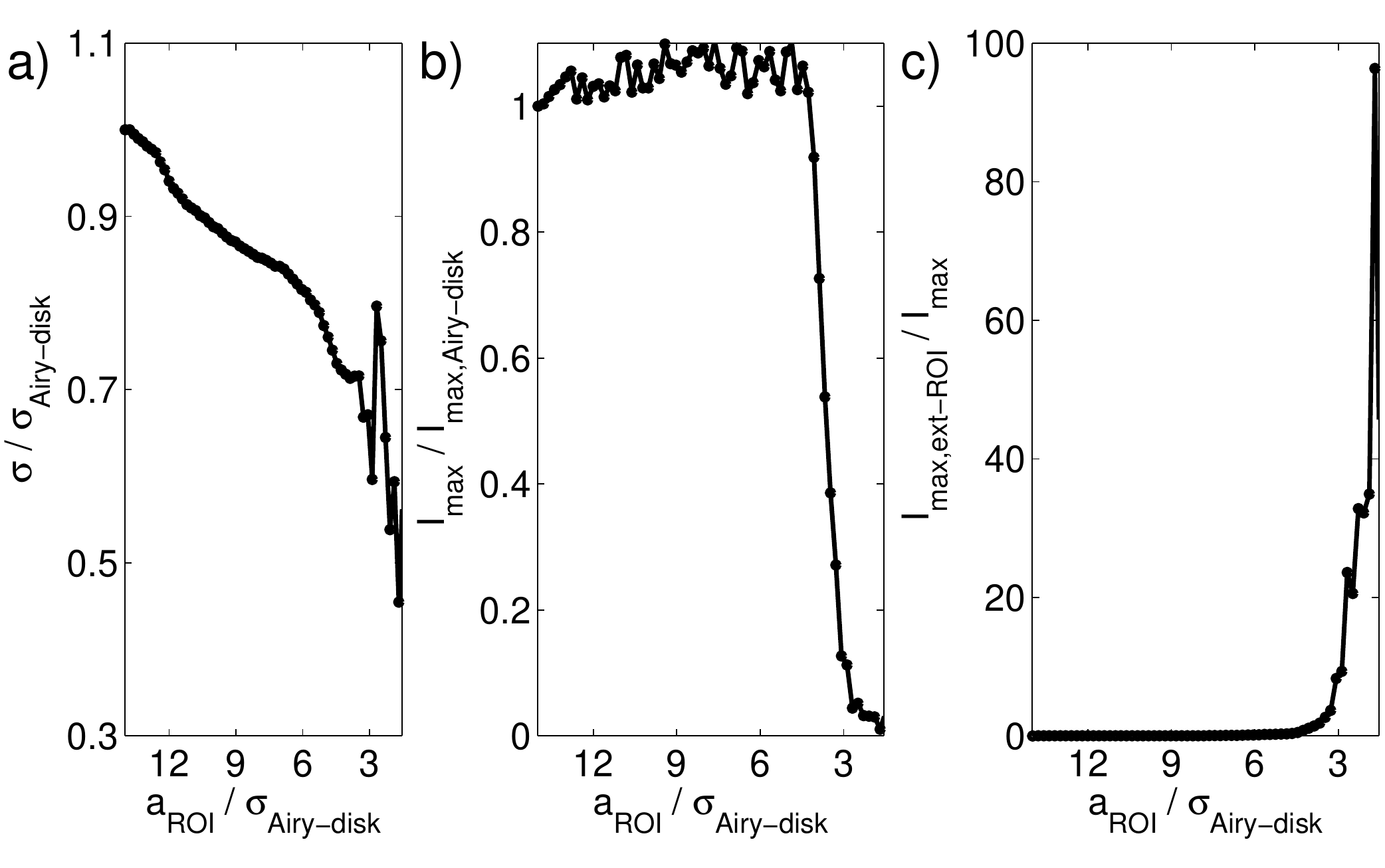}
\caption{ Quantitative analysis of the intensity profiles shown in Fig. \ref{fig:col2Dexp} as a function of the width $a_{\textrm{ROI}}$ of the ROI. (a) Width of the central spot $\sigma$ as defined by (\ref{eq:QMESSO2}). (b) Normalised maximum intensity $I_{\textrm{max}}$.  (c) Contrast ratio between the maximum intensity $I_{\textrm{max,ext-ROI}}$ outside of the ROI and maximum intensity $I_{\textrm{max}}$ of the central spot. All lengths are measured in units of $\sigma_{\textrm{Airy-disk}}=63\ \upmu\textrm{m}$.}
\label{fig:spotsize}
\end{figure}

\begin{figure}[tb]
\centering\includegraphics[width=8cm]{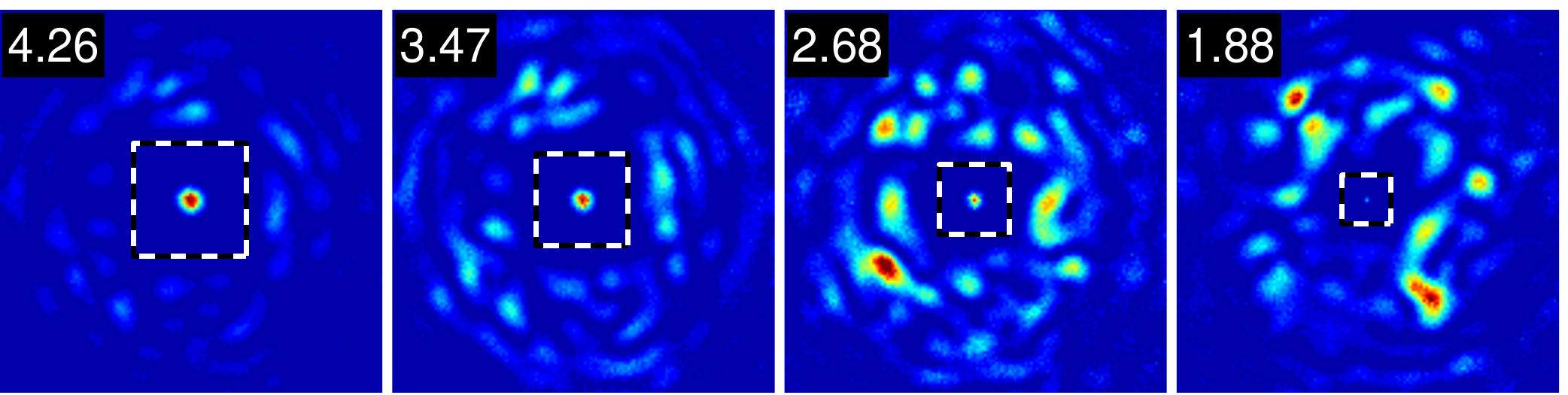}
\caption{2D intensity profiles obtained by numerically superimposing the measured fields $E_{u}(x,y,z_{T})$ in the target plane as opposed to the profiles shown in Fig.~\ref{fig:col2Dexp} obtained by SLM encoding the final superposition in the initial plane. Figure specifications are equivalent to Fig.~\ref{fig:col2Dexp}.}
\label{fig:figure3}
\end{figure}

We have also performed a quantitative analysis of the qualitative profiles shown in Fig.~\ref{fig:col2Dexp} which was based on the following parameters: 1. The central spot size $\sigma$ (determined as the FWHM of a Gaussian the central spot is fitted to) in units of $\sigma_{\textrm{Airy-disk}}$, 2. the peak intensity $I_{\max}$ of the central spot relative to the peak intensity $I_{\textrm{max,Airy-disk}}$ of the Airy disk and 3. the peak intensity $I_{\textrm{ext-ROI}}$ outside of the ROI relative to the central spot peak intensity $I_{\max}$. The respective parameters were plotted versus $a_{\textrm{ROI}}/\sigma_{\textrm{Airy-disk}}$ as shown in Fig.~\ref{fig:spotsize}(a)-(c). The plotting range extends to $a_{\textrm{ROI}}/\sigma_{\textrm{Airy-disk}}\approx2$ where the measured data became very noisy since the limits in terms of both SLM encoding and detector sensitivity were reached. All three parameters exhibit distinct changes when $\sigma/\sigma_{\textrm{Airy-disk}}\approx5$: 1. The relative spot size $\sigma/\sigma_{\textrm{Airy-disk}}$ fastly drops to a value below $0.5$ after having persistently decreased to approximately $0.7$ (see Fig.~\ref{fig:spotsize}(a)). 2. The central spot relative peak intensity $I_{\max}/I_{\textrm{max,Airy-disk}}$ manifests a sharp decrease of the relative eigenspot peak intensity (see Fig.~\ref{fig:spotsize}(b)). 3. The redistributed relative intensity $I_{\textrm{max,ext-ROI}}/I_{\textrm{max}}$ simultaneously experiences a vast increase (see Fig.~\ref{fig:spotsize}(c)). 

Overall, our results manifest remarkable similarities to the concept of superoscillations in band limited functions \cite{Zheludev2008}. The central spot size is decreased below the diffraction limit at the expense of the spot intensity which is redistributed to the so-called side bands around the spot. It is important to remark upon the difference and commonality between superoscillation and quadratic measure eigenmodes. Superoscillations relates to generation of spectrally bandwidth limited fields that produces oscillations with frequencies outside the limiting bandwidth. As such, super oscillating fields are based on the Fourier relationship between reciprocal space and real space. In contrast, our quadratic measure eigenmodes approach is a generic method that does not rely on Fourier relationship but can be used to optimise any general quadratic measure operator. The application of QME corresponds to minimising the spot size of a superposition of bandwidth limited fields and in this specific case, the QME is equivalent to super oscillation.

Although we are currently limited to minimum spot sizes of $\sigma_{\textrm{Airy-disk}}\approx0.5$ due to phase-only SLM encoding, limited SLM resolution, and limited detector resolution and sensitivity, our concept offers a huge potential. To demonstrate this we have performed a numerical superposition of the measured fields $E_{u}(x,y,z_{T})$ in the image plane. The resulting intensity distribution $I_{u}(x,y,z_{T})\propto|\sum_{u}E_{u}(x,y,z_{T})|^{2}$ is shown in Fig.~\ref{fig:figure3} and would be achieved with an advanced experimental configuration in particular featuring high resolution amplitude and phase SLM encoding to realize nearly-perfect linearity in our optical system. The central spot in the right graph has a size of $\sigma_{\textrm{Airy-disk}}\approx0.1$ that is the Airy disk is beaten by an order of magnitude.  

In this letter, we have experimentally used the QME approach to locally generate  sub-diffraction light spots and we demonstrate beating the diffraction limit by a factor of 4 in area. The theory that we employ is rigorous and based on considering the spot size operator as a quadratic measure originating from Maxwell's equations. Excitingly, we can define other quadratic measure operators to which our approach is applicable (see Table \ref{tab}). The generic nature of our approach means that it may be applied to optimise the size and contrast of optical dark vortices, the Raman scattering or fluorescence of any samples, the optical dipole force and the angular/linear momentum transfer in optical manipulation. Our approach is applicable to all linear physical phenomena where generalised fields interfere to give rise to quadratic measures.

\bibliographystyle{apsrev}
\bibliography{subdif}

\end{document}